# The Application of Green GDP and Its Impact on Global Economy and Environment: Analysis of GGDP based on SEEA model

Mingpu Ma[*]

[*] College of Management and Economics, Tianjin University, Tianjin 300072, China
E-mail: m125_0409@tju.edu.cn

**Abstract:** This paper presents an analysis of Green Gross Domestic Product (GGDP) using the System of Environmental-Economic Accounting (SEEA) model to evaluate its impact on global climate mitigation and economic health. GGDP is proposed as a superior measure to traditional GDP by incorporating natural resource consumption, environmental pollution control, and degradation factors. The study develops a GGDP model and employs grey correlation analysis and grey prediction models to assess its relationship with these factors. Key findings demonstrate that replacing GDP with GGDP can positively influence climate change, particularly in reducing $CO_2$ emissions and stabilizing global temperatures. The analysis further explores the implications of GGDP adoption across developed and developing countries, with specific predictions for China and the United States. The results indicate a potential increase in economic levels for developing countries, while developed nations may experience a decrease. Additionally, the shift to GGDP is shown to significantly reduce natural resource depletion and population growth rates in the United States, suggesting broader environmental and economic benefits. This paper highlights the universal applicability of the GGDP model and its potential to enhance environmental and economic policies globally.

# Contents

# 1 Introduction

## 1.1 Background

At present, global environmental problems are becoming more and more intense. Rising surface temperatures, melting glaciers, and rising sea levels warn us of the urgency and importance of protecting the environment. However, indicators to measure environmental issues are few and far between and not widely used. Countries must rely on something other than a constant hand to maintain their momentum for environmental protection.

Most countries use the gross domestic product (GDP) indicator to measure the health of the national economy. However, this indicator only considers current production relations and economic development, but not its impact on the environment and the country's ability to develop sustainably. As a result, this indicator needs to be improved in evaluating the proper economic health of the nations.

Green GDP (GGDP) can eliminate this flaw very well. If countries changed how they assess and compare their economies to GGDP, the world's climate crisis would be significantly mitigated.

Considering the background information and restricted conditions identified in the problem statement, we need to solve the following problems:

- Due to the increasing impact of GGDP on the world, many studies on the way to account for GGDP have emerged today. We need to choose one of the many methods and show the relationship between GGDP and the factors that constitute its correlation.

- The value of GGDP is calculated, and the indicator to measure global environmental impact is also selected. A model is constructed using the above calculations and hands to determine the relationship between GGDP and global environmental impact.

- Getting countries to replace GDP with GGDP will encounter many difficulties. A model is constructed to show that such a substitution is necessary, describing the advantages and disadvantages of replacing GDP with GGDP.

- Choose a specific country and measure the impact of converting GGDP to GDP for that country.

- Write a non-technical report to the leaders of the above countries asking for clarification and justification on whether they support a shift from GGDP to GDP.

## 1.2 Literature Review

Since the 1970s, the United Nations, national governments, and renowned scholars have been making painstaking efforts to build a green national economic accounting system with "green GDP" as the core. In 1971, the Massachusetts Institute of Technology (MIT) proposed the "Ecological Demand Indicator (ERI)" in an attempt to quantify the relationship between eco-nomic growth and the environment[1]. In 1972, Tobin and Nordhaus proposed the Net Economic Welfare Indicator (NEW), which advocates that the social costs of environmental pollution and other economic actions should be deducted from the GDP[2]. In 1990, World Bank economists Daly and Cobb proposed the Indicator of Sustainable Economic Welfare (ISEW), suggesting that social costs such as health care expenditures should not be counted as a contribution to the economy.[3]

Regarding practical exploration, Norway was the first country to start accounting for natural resources. Subsequently, Finland, France, Japan and other countries have designed national economic accounting frameworks that include resource and environmental accounting. Among them, *the Handbook on Integrated Environmental and Economic Accounting (SEEA1993)*[4], officially published in 1994, was widely accepted by countries worldwide. The manual put forward the basic framework of economic and environmental accounting and the concept of green GDP. The international research and practice of SEEA (1993), SEEA (2000), SEEA (2003) and **SEEA (2012)** were further published after a thorough summary and revision[5].

## 1.3 Our Work

By exploring the background of the problem and reanalyzing it, we have organized the ideas of this paper in the form of a fishbone diagram as follows.

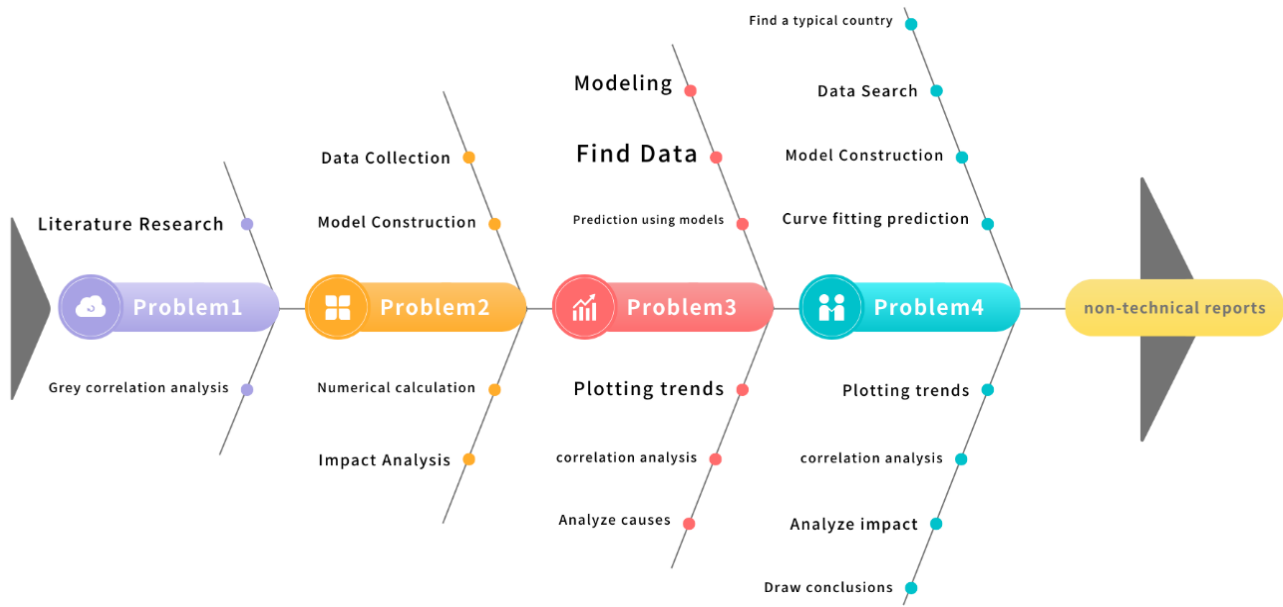

Figure 1: The Flow Chart of Our Work

# 2 Assumptions and Glossaries

To simplify the above problem, we follow the following assumptions in the problem-solving process.

Assumption 1: Only $CO_2$ emissions and surface temperature will significantly impact climate.

Justification: We know from some primary scientific connections that have been proven: that the concentration of greenhouse gases in the Earth's atmosphere directly affects the global average temperature; that the concentration of greenhouse gases has continued to rise since the time of the industrial revolution and that the global average temperature has increased; and that the most abundant greenhouse gas in the atmosphere, carbon dioxide, which accounts for about two-thirds of its total, is produced primarily by the burning of fossil fuels. So, we choose to measure the global climate impact in terms of both $CO_2$ emissions and surface temperature.

Assumption 2: Only the cost of urban environmental infrastructure invested by the government is considered in the accounting of the amount of value lost to environmental pollution control.

Justification: The cost of environmental pollution control includes the following three charges: the investment cost of the urban environment, the investment cost of industrial pollution sources, and the investment cost of construction of "three simultaneous" ecological protection[6]. However, because the investment cost of industrial pollution source treatment and the investment cost of construction of "three simultaneous" environmental protection belongs to the part of enterprise investment in medicine, which has been deducted as the production cost of enterprises, so only the cost of urban ecological invested by the government is considered in the process of GGDP construction.

Assumption 3: The CPI is the main factor affecting each country's economy.

Justification: The Consumer Price Index (CPI) is a critical macroeconomic indicator reflecting the changes in the price level of consumer goods and services related to people's lives and is also an important indicator for macroeconomic analysis and decision-making as well as national economic accounting. CPI influences the introduction and strength of national macroeconomic control measures and provides an understanding of price changes worldwide. Therefore, this paper uses CPI and employment rate as the primary research indicators of the economy.

The necessary mathematical notations used in this paper are listed in Table 1.

**Table 1: Glossaries used in this paper**

| Glossary | Meaning |
|---|---|
| GGDP | Green Gross Domestic Product |
| GDP | Gross Domestic Product |
| RDM | Resource Depletion Minus Value |
| EPCL | Value of Environmental Pollution Control Losses |
| EPDL | Value of Environmental Pollution Degradation Loss |

# 3 Select the Method of Calculating GGDP

In this section, we construct a specific method for calculating GGDP concerning the published findings. Also, some reliable indications are used to analyze the impact of each indicator on climate.

## 3.1 Data Collection

To ensure the comprehensiveness and authority of our data, we choose the following websites as our data sources.

Table 2: Data Sources

| Data Sources | Website |
| --- | --- |
| UNdata | http://data.un.org/Default.aspx |
| World Bank Open Data | https://data.worldbank.org/ |
| World Environment Situation Room | https://wesr.unep.org/ |

To ensure the scientific accuracy of our analysis, we select a panel of global data spanning from 1990 to 2020.

## 3.2 Selection of SEEA model

The accounting methods used for GGDP accounting are divided into two main categories; one is a direct measurement based on the same principles as traditional GDP accounting and includes two methods: the production method and the expenditure method. The production method, also is the total output obtained by deducting intermediate inputs from the work of each industrial sector, except that the intermediate inputs for accounting GGDP include not only the economic assets consumed, but also the natural assets. In the expenditure approach, the output of natural resource depletion is added to the original accounting. Another type of accounting method is indirect accounting, which is based on traditional GDP accounting and adds resource, environmental, social and other indicators to obtain the GGDP value. Indirect accounting can be divided into SEEA-based accounting, external economy and external diseconomy, social welfare, input-output, material flow analysis, energy value analysis, etc.

At present, countries generally adopt the United Nations SEEA accounting system as the framework for green GDP accounting, so this paper also takes SEEA (2012) as the basis for green GDP accounting. It considers the cost of natural resource depletion and environmental loss as deductions in light of the actual situation of global economic and social development[7].

The specific GGDP accounting formula is:

$$GGDP = GDP - RDM - EPCL - EPDL \qquad (1)$$

In this paper, according to the commonality of domestic and foreign resource classification markers and the SEEA accounting framework system, natural resources are classified into the water, energy, arable land, forest, and fishery resources. The value of environmental pollution control loss includes actual control investment and virtual control costs. The value of ecological pollution degradation loss mainly consists of the cost of human health damage caused by environmental pollution degradation. The second is the value of ecological pollution degradation loss, which primarily includes the cost of human health damage caused by environmental pollution degradation and the failure of accelerated depreciation of fixed assets. The second is the value of ecological pollution degradation loss, which mainly includes the cost of human health damage caused by environmental pollution degradation, the cost of accelerated depreciation of fixed assets and the cost of natural disasters.

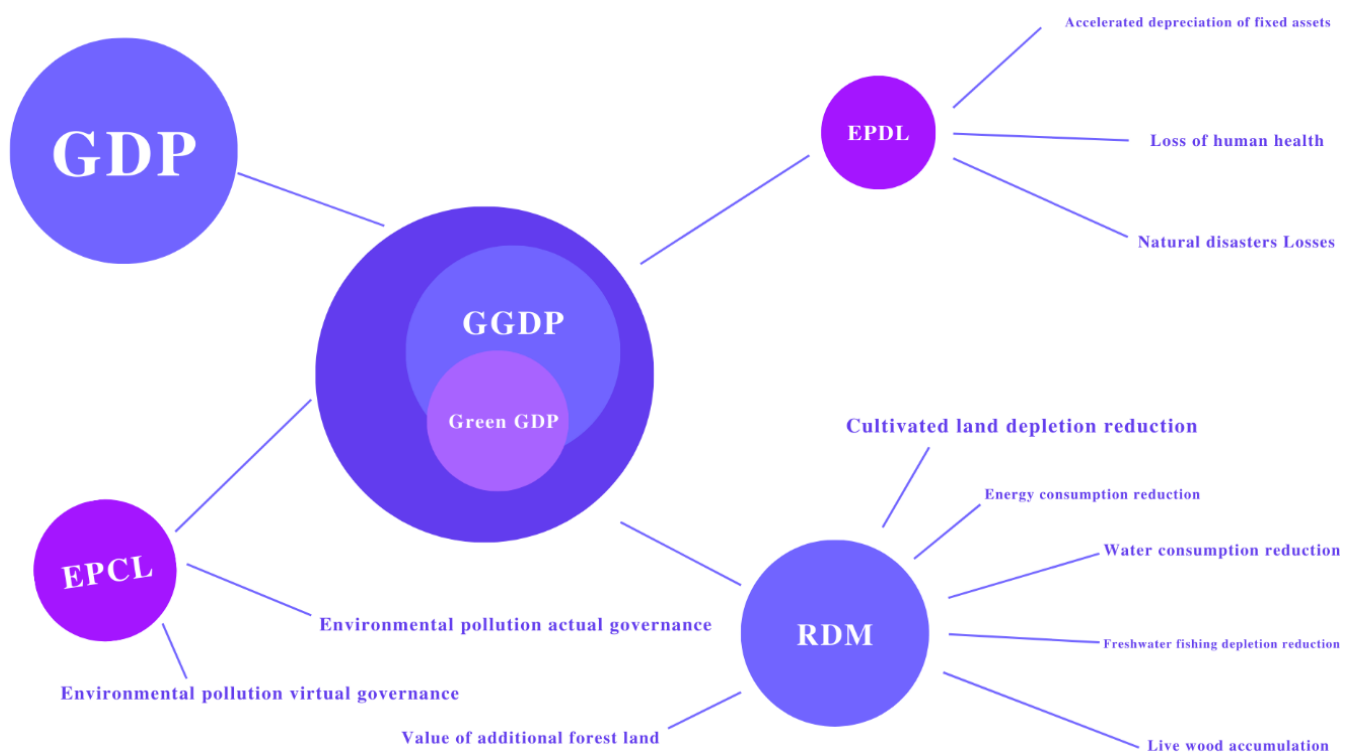

Figure 2: Secondary indicators corresponding to the GGDP calculation

## 3.3 Calculation by grey correlation analysis

The gray correlation analysis model (GRA) calculates the relationship between GGDP and the four level 1 indicators under it, and subsequently the relationship between the four level 1 indicators and their respective level 2 indicators. This is used to determine the impact of GGDP on global climate. The specific steps of the solution are as follows:

**Step 1: The variables were preprocessed to remove the effect of magnitude, and the range of variables was narrowed to simplify the calculation.**

Let the normalized matrix be $Z$, and the elements in are denoted as $Z_{ij}$:

$$Z_{ij} = \frac{x_{ij}}{\overline{x}_{ij}} \qquad (2)$$

The normalized matrix Z is obtained:

$$Z = \begin{bmatrix} z_{11} & z_{12} & \cdots & z_{1m} \\ z_{21} & z_{22} & \cdots & z_{2m} \\ \vdots & \vdots & \ddots & \vdots \\ z_{n1} & z_{n2} & \cdots & z_{nm} \end{bmatrix} \qquad (3)$$

**Step 2: Define parent and child sequences.**

Parent series (also known as reference series): data series that reflects the system's characteristics, similar to the dependent variable $Y$, here noted as $x_0$.

Subseries (also called comparison series, compare series): a data series of factors affecting the system's behavior, which is similar to the independent variable $X$, here denoted as $(x_1, x_2, \cdots, x_m)$.

**Step 3: Calculate the correlation coefficients of each indicator in the subseries with the parent series.**

Define the gray coefficient, i.e., the correlation coefficient of each indicator, as:

$$y(x_0(k), x_i(k)) = \frac{a + \rho b}{|x_0(k) - x_i(k)| + \rho b} \quad (i=1,2,\cdots,m,\ k=1,2,\cdots,n) \qquad (4)$$

In the above equation, $a$ is the minimum difference between the two poles, $b$ is the maximum difference between the two poles, and $\rho$ is the resolution factor. The value of this paper is 0.5.

$$a = \min_i \min_k |x_0(k) - x_i(k)| \qquad (5)$$

$$b = \max_i \max_k |x_0(k) - x_i(k)| \qquad (6)$$

Firstly, the intermediate difference matrix is obtained according to the above steps. The two-level minimum and maximum differences are accepted according to the matrix. Finally, the correlation coefficient matrix is calculated according to the correlation coefficient formula.

**Step 4: Calculate the grey correlation and conclude.**

Define the gray correlation, i.e., the mean value of each column of the correlation coefficient matrix.

$$y(x_0, x_i) = \frac{1}{n} \sum_{k=1}^{n} y(x_0(k), x_i(k)) \quad (7)$$

The final grey correlations of GGDP with the primary indicators are shown in the table below.

Table 3: Gray correlation between GGDP and primary indicators

| Indicators | Relevance Index |
|---|---|
| GDP | 0.9247 |
| RDM | 0.5901 |
| EPCL | 0.5365 |
| EPDL | 0.5353 |

The correlation coefficients between each primary indicator and its corresponding secondary indicator are shown in the table below.

Table 4: Correlation coefficient between RDM and its secondary indicators

| Indicators | Relevance Index |
|---|---|
| Cultivated land depletion reduction | 0.4294 |
| Energy consumption reduction | 0.9087 |
| Water consumption reduction | 0.4271 |
| Freshwater fishing depletion reduction | 0.4307 |
| Live wood accumulation | 0.4345 |
| Value of additional forest land | 0.4330 |

Table 5: Correlation coefficient between EPCL and its secondary indicators

| Indicators | Relevance Index |
|---|---|
| Environmental pollution actual governance | 0.9236 |
| Environmental pollution virtual governance | 0.5533 |

Table 6: Correlation coefficient between EPDL and its secondary indicators

| Indicators | Relevance Index |
|---|---|
| Accelerated depreciation of fixed assets | 0.4677 |
| Loss of human health | 0.9834 |
| Natural disasters Losses | 0.4727 |

From the above analysis, we can see that GGDP is most correlated with GDP, but RDM, EPLC, and EPDL will also significantly impact it. Meanwhile, the most significant impact on RDM is *Energy consumption reduction*, the most considerable effect on EPCL is *Environmental pollution actual governance*, and the most significant impact on EPDL is *Loss of human health*. Therefore, replacing GDP with GGDP as a measure of the economy will positively affect climate change, especially the three factors mentioned above.

The comparison of GGDP and GDP calculated based on the above analysis is shown in the figure below.

# 4 Building GGDP & Analyzing the Impact on Climate

## 4.1 Data processing & Model Building

From the correlation above analysis, it is clear that we can choose the most significant secondary indicators to replace the three critical factors affecting GGDP.

Meanwhile, for the sake of calculation, we take RDM, the most crucial first-order indicator besides GDP, as the main object of study. The EPCL is fitted to the EPDL as a function of the data.

The relationship between the values of RDM and EPCL is shown in the following figure.

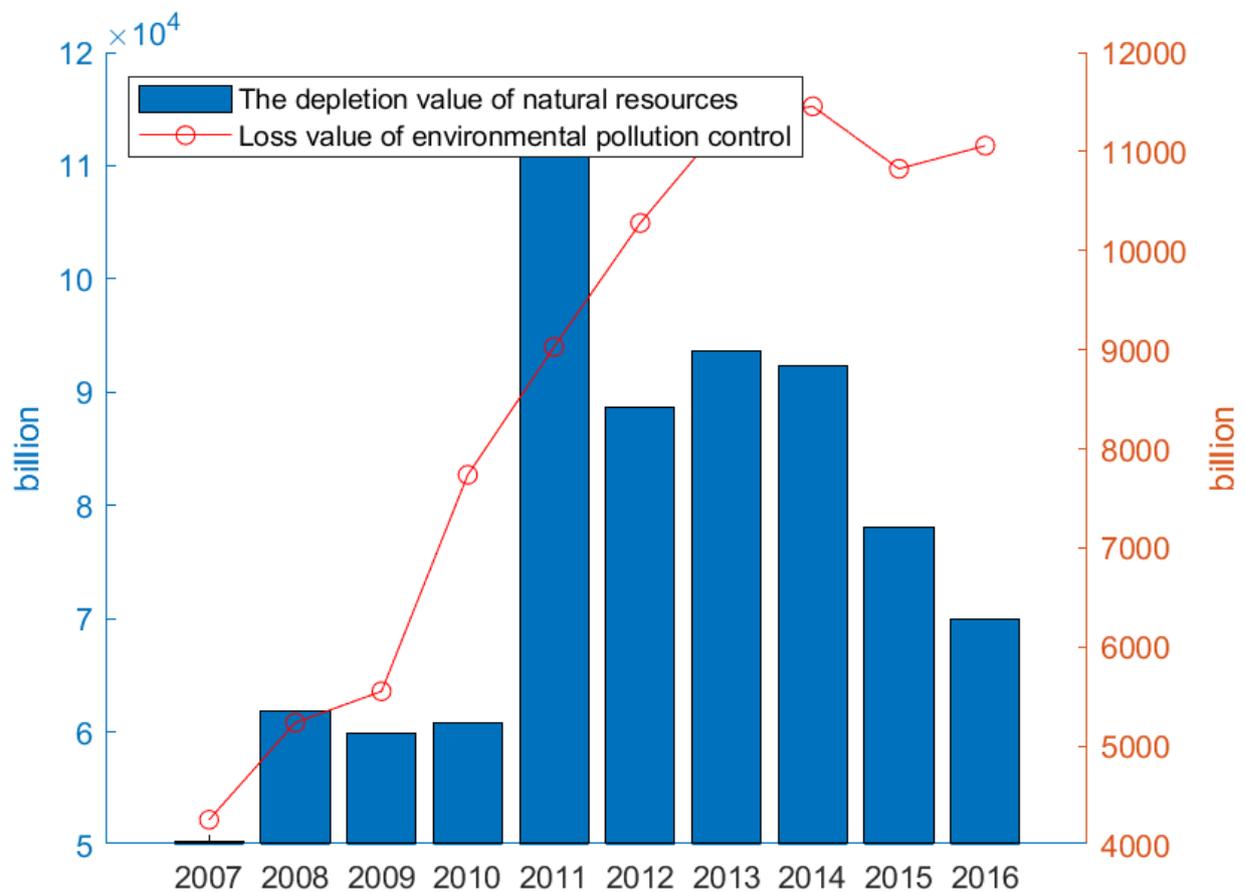

Figure 3: Numerical relationship between RDM and EPCL

Since the trend between the two is relatively close, we fit the relationship between them with the following fitting equation:

$$y = 0.1009x + 932.2 \qquad (8)$$

The numerical relationship between RDM and EPDL is shown in the following figure.

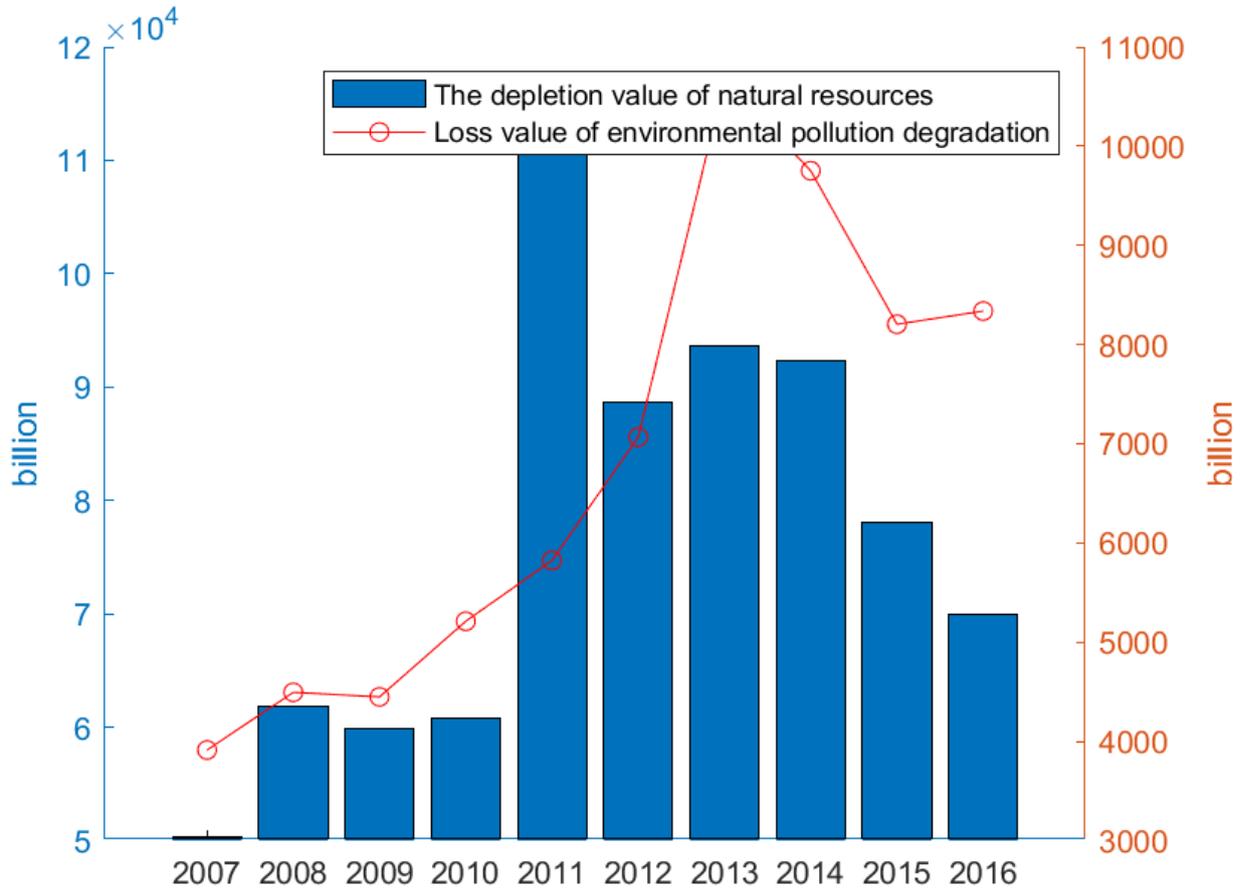

**Figure 4: Numerical relationship between RDM and EPDL**

The functional relationship between the two is shown below:

$$y = 0.07316x + 1179 \qquad (9)$$

We can calculate the specific value of RDM using the relationship between gross national income (GNI) and its percentage of RDM. Defining this percentage $\delta$, the formula to calculate the RDM is as follows:

$$RDM = \delta \times GNI \qquad (10)$$

Based on the functions above, we can calculate the specific values of GGDP by combining the particular importance of the GDP and RDM of each country. Five typical representatives of developed and developing countries are selected below: China, the United States, India, Spain, and Germany. The values are plotted as follows.

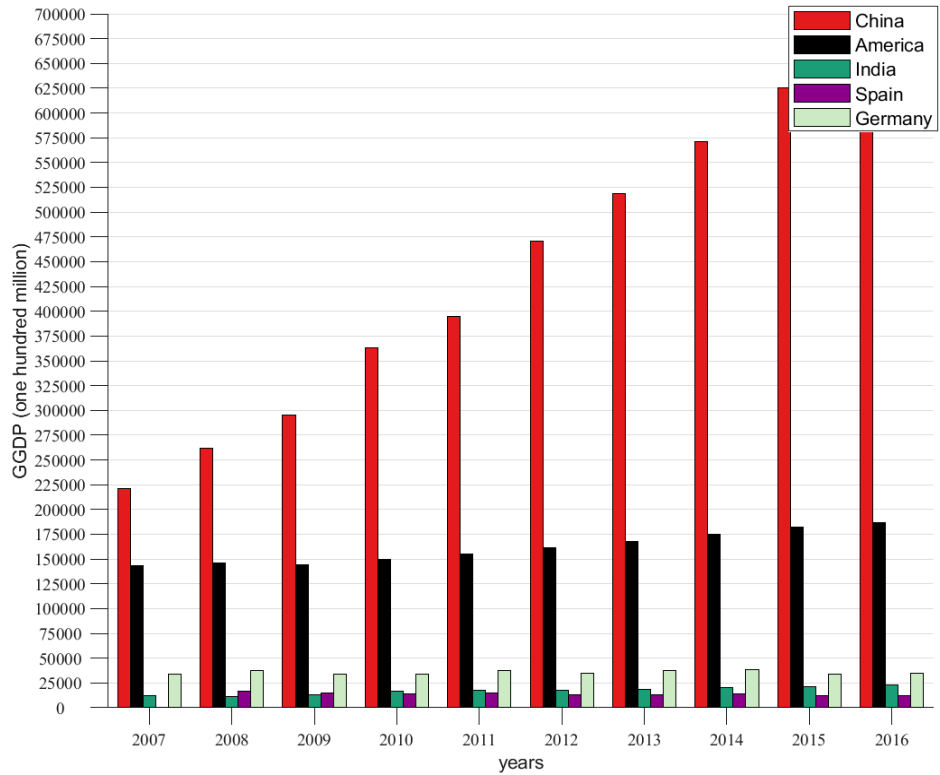

**Figure 5: Values of GGDP for five typical countries**

## 4.2 Analysis of the impact on climate

We use the degree of change in surface temperature and $CO_2$ emissions as an important indicator of climate. The closer the degree of change is to zero, the more significant the environmental impact of using GGDP instead of GDP can be considered.

The chart between surface temperature, $CO_2$ emissions and GGDP is as follows:

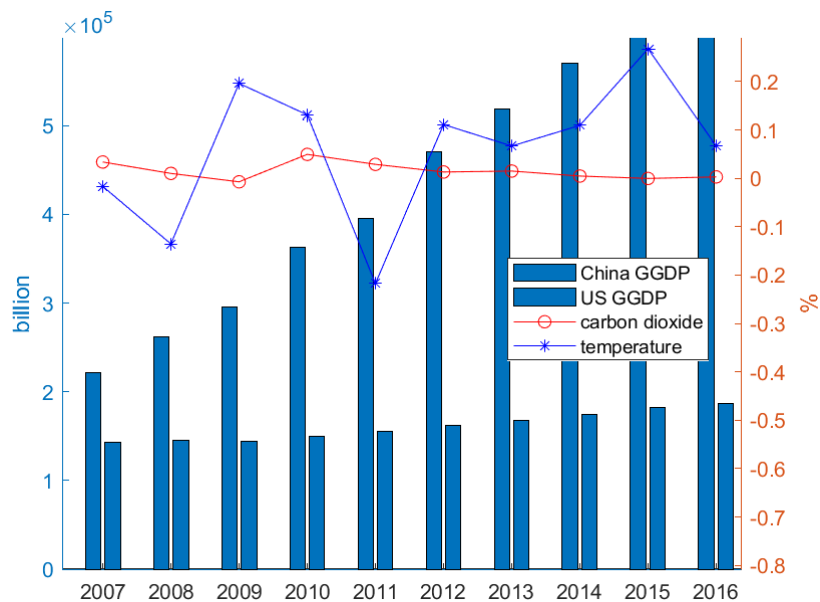

**Figure 6: Chart between surface temperature, $CO_2$ emissions and GGDP**

The above image shows that the $CO_2$ emissions are close to zero and the amount of change in surface temperature fluctuates between plus and minus 0.2%. Thus, it can be concluded that using GGDP instead of GDP significantly impacts the environment.

## 5 Forecasting the Impact of GGDP on the Economy

### 5.1 Building predictive models

By using gray forecasting, we can predict the future trend of GGDP and other values for each country in the world. The specific steps of gray forecasting are as follows.

**Step 1: Construct the first order differential equation for GM (1, 1).**

$$\frac{dx}{dt} + ax = u \tag{11}$$

**Step 2: From the definition of the derivative** $\frac{dx}{dt} = \lim_{\triangle t \to 0} \frac{x(t+\triangle t) - x(t)}{\triangle t}$ **, when** $\triangle t$ **is infinitely small, the approximation has the following discrete form.**

$$\frac{\triangle x}{\triangle t} = x(k+1) - x(k) = \triangle^{(1)}(x(k+1)) \tag{12}$$

**Step 3: Replace** $x^{(i)}(i)$ **by the average of the two moments before and after.**

**Step 4: Organize the above equations and write them as matrix expressions.**

$$\begin{bmatrix} x^{(0)}(2) \\ x^{(0)}(3) \\ M \\ x^{(0)}(N) \end{bmatrix} = \begin{bmatrix} -\frac{1}{2}[x^{(1)}(2) + x^{(1)}(1)] & 1 \\ -\frac{1}{2}[x^{(1)}(3) + x^{(1)}(2)] & 1 \\ M & 1 \\ -\frac{1}{2}[x^{(1)}(N) + x^{(1)}(N-1)] & 1 \end{bmatrix} \begin{bmatrix} a \\ u \end{bmatrix} \tag{13}$$

**Step 5: Derive the resulting equation.**

$$\hat{x}^{(1)}(k+1) = \left[x^{(1)}(1) - \frac{\hat{u}}{\hat{a}}\right] e^{-\hat{a}k} + \frac{\hat{u}}{\hat{a}} \tag{14}$$

### 5.2 Analysis results

With the above steps, we can predict the trend of GGDP and consumption capacity in each country. Here, we use China as a typical developing country and the United States as a specific developed country to forecast their GGDP and the leading indices affecting their consumption.

The projections for China's GGDP and CPI are shown below.

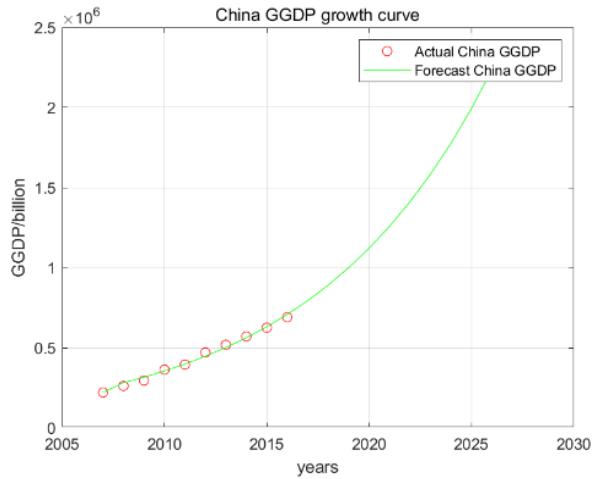 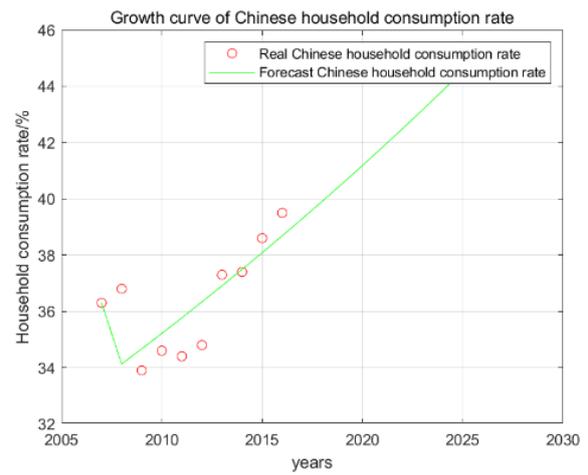

(a) China GGDP Forecast Curve      (b) China CPI Forecast Curve

**Figure 7: Growth curve of related projects in China**

The projections for GGDP and CPI of the United States are shown below.

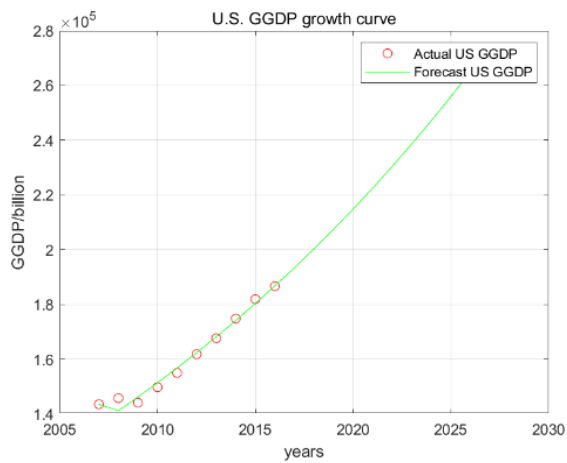 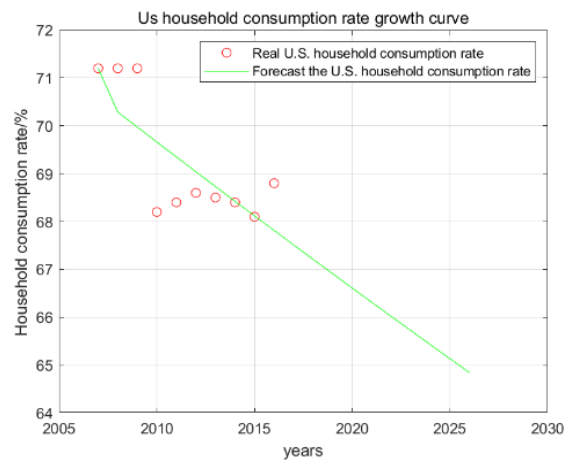

(a) U.S. GGDP Forecast Curve      (b) U.S. CPI Forecast Curve

**Figure 8: Growth curve of related projects in the United States**

According to the correlation analysis, it can be seen that the correlation coefficients of the upward or downward trend of GGDP and CPI in China and the United States are 0.7481 and 0.4910, which are well correlated. This can indicate that the change in consumption level arises with the process of using GGDP instead of GDP.

**Based on the above results, we can conclude the following:**

1. for developing countries represented by China, using GGDP instead of GDP will increase the economic level.

2. For developed countries represented by the United States, using GGDP instead of GDP will make the economic level fall.

3. The possible reason for this change is that using GGDP instead of GDP will decrease people's living standards. Still, because Chinese people have better environmental awareness, it can offset this effect and even cause the CPI indicator to increase.

4. Since the correlation coefficient between GGDP and CPI does not reach 0.8, they are only moderately correlated. So, this situation may occur due to other factors.

# 6 Exploring the specific impact in a typical country

## 6.1 Build the model and analyze the results

Because the U.S. is typical among developed countries and considering the ease of data collection, this paper uses the U.S. as a typical representative for analysis.

Utilizing curve-fitting projections, we project total U.S. natural resource consumption against the population growth rate, as shown in the following figure.

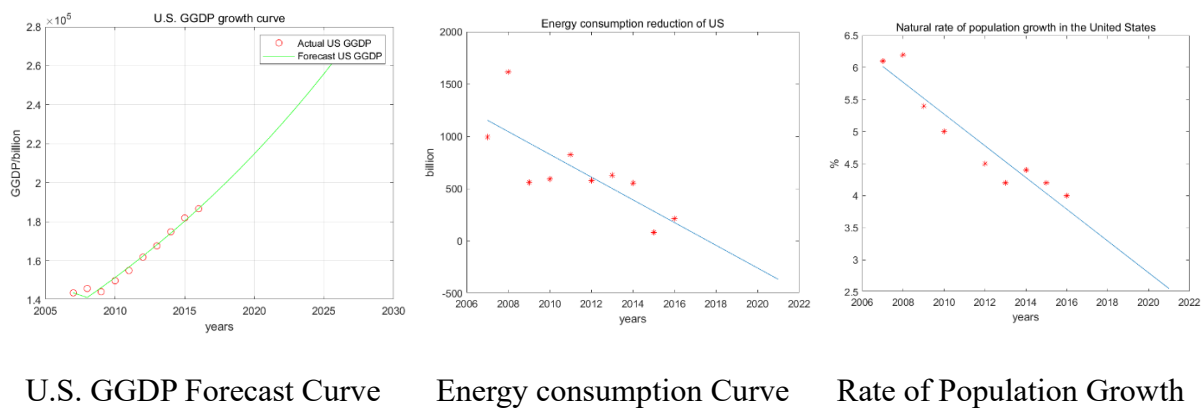

    U.S. GGDP Forecast Curve    Energy consumption Curve    Rate of Population Growth

**Figure 9: Forecast curves for U.S. correlates**

Based on the correlation analysis, we can conclude that the correlation coefficients of GGDP change with population change and natural resource depletion change in the U.S. are 0.4305 and 0.9329, respectively, thus indicating that GGDP change is strongly correlated with natural resource depletion change and moderately correlated with population growth rate.

## 6.2 Summarize the impact of the shift on the United States

The above analysis shows that natural resource depletion will be significantly reduced when the U.S. chooses GGDP instead of GDP. This reduction is strongly correlated with GGDP.

This reduction in natural resource depletion in the coming years implies higher energy use in the US. More importantly, reducing natural resource depletion means that traditional heavy industries will emit fewer emissions, increasing industrial efficiency and improving the environment.

At the same time, when GGDP replaces GDP, the natural population growth rate in the U.S. will decline. Although this change is not strongly correlated with GGDP, it is still possible to analyze the impact of GGDP on population growth.

A decrease in the population growth means the pressure on the U.S. to raise children decreases. At the same time, as a result of the shift from GDP to GGDP, the per capita environmental capacity of the U.S. will increase, thus making Americans more environmentally conscious in general.

Putting the two together, a decline in natural resource depletion means increased natural resource utilization. In contrast, reducing the population growth rate makes supporting future generations less of a burden. Thus, replacing GDP with GGDP has many benefits for the United States.

# 7 Model Evaluation and Further Discussion

## 7.1 Strengths

- This paper adopts the most advanced model in the field of green GDP, namely the SEEA (2012) model, to make our research conclusions more reliable.

- In this model, the function relation is obtained through fitting to make the simplified operation as accurate as possible.

- This model finds the stock indexes of climate impact and consumption, which makes calculating various countries' mate impact and consumption habits more accurate.

- Based on the model in this paper, we draw several authentic trend images to make the model more fully expressed.

## 7.2 Weaknesses

- Many detailed second-level indicators are difficult to obtain, so that this model may have errors with the actual situation.

- Due to technical limitations, the model does not show all the factors of green GDP.

## 7.3 Further Discussion

- We can explore the values of all secondary indicators more clearly, so that the importance of GGDP will be more accurate.

- We can use more detailed data to give a clearer picture of climate impact in question two and consumption power in question three.

- By reading more literature, we can extend this model to more areas of environmental impact and make it more useful.

# 8 Conclusion

In summary, our team adjusted the SEEA method to measure GGDP. After accounting for the data support, we found that among the primary indicators measured by GGDP, GGDP has the strongest correlation with GDP. Still, RDM, EPLC and EPDL also have significant effects on it. Meanwhile, after analyzing the correlation between primary and secondary indicators, we found that energy consumption reduction has the most substantial impact on RDM, actual environmental pollution control has a considerable impact on EPCL, and human health loss has the most significant effect on EPDL. There is also a correlation between these secondary indicators and the global climate. In the global environmental analysis, we selected the degree of change in surface temperature and carbon dioxide emissions as indicators of the environment. Through the research, we found that using GGDP instead of GDP as a measure of the economy would cause the change in both of these to converge to zero and positively impact the global climate. After analyzing the impact of GGDP on environment, we examined the effect of replacing GDP with GGDP on the economic consumption and living standards of each country in the world. We selected China and the United States to represent developing and developed countries. We found that this substitution would increase developing countries' economic levels and decrease developed countries' financial status. In exploring the impact of replacing GDP with GGDP on countries, we chose the United States for our analysis. We found that replacing GDP with GGDP would significantly reduce the depletion of natural resources in the U.S. and reduce the natural growth rate of the population by a certain amount; combining the two, we believe that replacing GDP with GGDP would bring many benefits to the U.S.

# Appendices

| **Appendix 1** |
|---|
| Introduce: Code used to calculate the relationship between GGDP and its corresponding level 1 indicators showing grey correlation analysis. |

```
clear;cc;
load data1.mat;
r = size(data1,1);
c = size(data1,2);
% In the first step, the variables are preprocessed to eliminate the effect of magnitude
%avg = repmat(mean(data),r,1);
%data = data./avg;
% Define parent and child sequences
Y = data1(:,1); % Parent sequence
X = data1(:,2:c); % Subsequences
Y2 = repmat(Y,1,c-1); % Copy the parent sequence to the right to column c-1
absXi_Y = abs(X-Y2)
a = min(min(absXi_Y)) % Global Minimum
b = max(max(absXi_Y)) % Global Maximum
ro = 0.5; % The resolution factor is 0.5
gamma = (a+ro*b)./(absXi_Y+ro*b) % Calculate the correlation coefficients of each indicator in the subseries with the parent series
disp('The gray correlations of each indicator in the subseries are:');
ans = mean(gamma)
```

| **Appendix 2** |
|---|
| Introduce: Taking China's CPI forecast as an example, it shows the concrete code of the grey forecast model. |

```
% Establish sign variables a(development coefficient) and b (grey action)
syms a b;
c = [a b]';
% Primitive sequence A
A = [36.3 36.8 33.9 34.6 34.4 34.8 37.3 37.4 38.6 39.5];
n = length(A);
%accumulation
B = cumsum(A);
for i = 2:n
C(i) = (B(i) + B(i - 1))/2;
end
C(1) = [];
B = [-C;ones(1,n-1)];
```

```
Y = A; Y(1) = []; Y = Y';
c = inv(B*B')*B*Y;
c = c';
a = c(1); b = c(2);
F = []; F(1) = A(1);
for i = 2:(n+10)
F(i) = (A(1)-b/a)/exp(a*(i-1))+ b/a;
end
G = []; G(1) = A(1);
for i = 2:(n+10)
G(i) = F(i) - F(i-1); % Get the predicted data
end
disp(' The forecast data is:');
G
H = G(1:10);
epsilon = A - H;
delta = abs(epsilon./A);
disp(' Relative residual Q test:')
Q = mean(delta)
% Draw a graph
t1 = 2007:2016;
t2 = 2007:2026;
plot(t1, A,'ro'); hold on;
plot(t2, G, 'g-');
xlabel('years'); ylabel('Household consumption rate/%');
legend('Real Chinese household consumption rate',' Forecast Chinese household consumption rate');
title('Growth curve of Chinese household consumption rate');
grid on;
```